\documentclass[a4paper]{article}
\usepackage{amssymb}
\usepackage{natbib}
\usepackage{hyperref}
\usepackage{graphicx}
\newcommand{\R}{\mathbb{R}}
\newcommand{\C}{\mathbb{C}}
\newcommand{\N}{\mathbb{N}}
\def\O{{\rm O}}
\def\A{{\rm A}}
\def\B{{\rm B}}
\def\d{{\rm d}}

\newcommand\maath{\mathsurround=0pt}
\newcommand{\EQM}[1]{\vcenter{\normalbaselines\maath
    \ialign{${\displaystyle ##}$\hfil&&\ ${\displaystyle ##}$\hfil\crcr
    \mathstrut\crcr\noalign{\kern-\baselineskip}
    \noalign{\smallskip}
#1\crcr\mathstrut\crcr\noalign{\kern-\baselineskip}}}}

\title{\large{\bf How many Keplerian arcs are there between two points of spacetime?}}

\author{}

\date{}

\begin{document}
	
	\maketitle
	
	\begin{center}
	{\bf Alain Albouy$^{1}$,\qquad Antonio J. Ure\~na$^{2}$}
	
	\bigskip
	$^{1}$ IMCCE, CNRS-UMR8028, Observatoire de Paris
	
        77, avenue Denfert-Rochereau, 75014 Paris, France
         
         Alain.Albouy@obspm.fr
         
         \bigskip
         
         $^{2}$ Departamento de Matematica Aplicada, Facultad de Ciencias
         
         Campus Universitario de Fuentenueva
         
         Universidad de Granada, 18071, Granada, Spain
         
        ajurena@ugr.es
        
        \bigskip
	\end{center}
	
{\bf Abstract.} We consider the Keplerian arcs around a fixed Newtonian center joining two prescribed distinct positions in a prescribed flight time. We prove that, putting aside the ``opposition case'' where infinitely many planes of motion are possible, there are at most two such arcs of each ``type''. There is a bilinear quantity that we call $b$ which is in all the cases a good parameter for the Keplerian arcs joining two distinct positions. The flight time satisfies a ``variational'' differential equation in $b$, and is a convex function of $b$.

\bigskip
\centerline{\bf 1. Introduction}
\bigskip

\citet{gauss} wrote in his Theoria motus: 

``Hence, inversely, it is apparent that two radius vectors given in magnitude and position, together with the time in which the heavenly body describes the intermediate space, determine the whole orbit. But this problem, to be considered among the most important in the theory of the motions of the heavenly bodies, is not so easily solved, since the expression of the time in terms of the elements is transcendental, and, moreover, very complicated.''

What Gauss calls ``this problem'' is called today Lambert's problem: {\it given a center of attraction, an initial position and a final position, find an arc of Keplerian orbit joining the two positions in a given time $\tau$}. Gauss gave two numerical methods to solve this problem, and modern authors have published dozens of efficient methods. Strangely enough, despite the prestigious mathematical descent of the problem, the basic question of the number of the solutions remained open. This number is correctly guessed for example in \citet{gooding}. We only know one publication addressing the question of the proof, by \citet{simo}, and giving a proof in the main cases. But a further convexity argument is needed to get this number in the multirevolution cases, requiring long computations which are only sketched by Sim\'o. We have already presented some simple proofs in \citet{AlU}. We can now complete this preliminary work. The unsolved questions about the number of the solutions completing a prescribed number of half-turns are now solved by observing a linear differential equation of the first order. We get stronger results by introducing two new parameters for the family of arcs with given ends. The flight time $\tau$ is a convex function of each of them. The first parameter, the inverse of the angular momentum $C$, becomes singular when the line from the initial to the final positions includes the attracting center. We will explain why this singularity is a necessary consequence of a good property: the energy is a rational function of $C$ or $1/C$. The second parameter $b$ does not have this singularity: it is a good parameter in all the cases. But the expression of the energy contains square roots.

{\it To solve Lambert's problem} is to compute the Keplerian arcs corresponding to a given triangle $\O\A\B$ and a positive flight time. Lambert's problem is a root finding problem, where the question is to find the values of $C$ or $b$ which correspond to the given value of $\tau$. Our convexity results have an important practical consequence: Newton's method for finding the roots is robust.

\citet{henon} found numerous cases where several Keplerian arcs start from a point of spacetime and arrive at another point of spacetime. In his study, one of the arcs is described by the Earth on its nearly circular orbit. He proposed a second arc as a possible trajectory for an interplanetary spacecraft, which would return to the Earth without spending energy.

Beyond the practical applications, a surprising aspect of our Theorems is their uniformity. Putting aside some obvious exceptions, the number of arcs does not depend on the initial and the final positions. It is the same whatever the prescribed number of completed turns. The arcs change continuously but never disappear when the prescribed flight time is increased.

\bigskip
\centerline{\bf 2. Statement of the main results}
\bigskip

{\bf 2.1. Definitions.} A {\it Keplerian arc} is a solution of the Kepler problem restricted to a finite interval of time. The length of this interval is called the {\it flight time}. The center of attraction is denoted by $\O$. The arc starts at a point $\A$ and ends at a point $\B$.  For a nonflat triangle $\O\A\B$ the {\it type} $k$ of a Keplerian arc is the number of completed half-turns around $\O$. An arc of type $k\geq 2$ is called {\it multirevolution}. The arc then belongs to an elliptic solution of the Kepler problem. If for example $k=2$ or $k=3$, the arc starts at $\A$, passes through $\B$ and $\A$, and ends at $\B$. If $k=0$ or $k=1$, the arc is called {\it simple} and the Keplerian solution may be elliptic, parabolic or hyperbolic.

{\bf 2.2. Theorem.}  Given any three points $\O$, $\A$ and $\B$ forming a nonflat triangle, and any flight time $\tau>0$, there is exactly one arc of type 0 and exactly one arc of type 1.
Given any integer number $k\geq 2$, there is a positive real number ${\cal T}_k$ such that the number of  arcs of type $k$ is zero if $\tau<{\cal T}_k$, one if $\tau={\cal T}_k$, two if $\tau>{\cal T}_k$.

{\bf Proofs.} See \S 6 for a proof and \S 8 for a second proof.

{\bf Euler's formula and related formulas.} Let $r_\A=\|\O\A\|$, $r_\B=\|\O\B\|$, $c=\|\A\B\|$, $s=r_\A+r_\B$. At the unique parabolic arc of type 0 and at the unique parabolic arc of type 1 the flight times are respectively
\begin{equation}\label{E*}
{\cal T}_0=\frac{(s+c)^{3/2}-(s-c)^{3/2}}{6},\quad {\cal T}_1=\frac{(s+c)^{3/2}+(s-c)^{3/2}}{6}.
\end{equation} 
Let $\Delta/2$ be the oriented area of the triangle $\O\A\B$. The respective angular momenta of these parabolic arcs are
\begin{equation}\label{C*}
{\cal C}_0=\frac{2\Delta}{\sqrt{s^2-c^2}(\sqrt{s+c}-\sqrt{s-c})},\qquad {\cal C}_1=-\frac{2\Delta}{\sqrt{s^2-c^2}(\sqrt{s+c}+\sqrt{s-c})}.
\end{equation}
Here Heron's formula for the area of the triangle
$4\Delta^2=(s^2-c^2)(c^2-(r_\A-r_\B)^2)$ gives a simplification and the value  ${\cal C}_0^{-2}={\cal C}_1^{-2}=(r_\A^{-1}+r_\B^{-1})/2$ in the case $s=c$.

{\bf Remark.} \citet{euler1} obtained the first formula (\ref{E*}) as the result of a simplification in the end of a computation. As it depends only on $c$ and $s$, ${\cal T}_0$ is the same for all the class of triangles with same $s$ and $c$, which includes a flat triangle where $\B$ is between $\O$ and $\A$. For such a triangle, the motion is rectilinear and formula (\ref{E*}) is easy to understand. In 1761, Lambert proved that the flight time, the energy, $s$ and $c$ are functionally dependent.  Euler immediately expressed to Lambert his admiration. Many proofs were subsequently proposed, as well as sign discussions (see \citealt{alb}). 

{\bf 2.3. Proposition.} For a nonflat triangle $\O\A\B$, the above special flight times satisfy $0<{\cal T}_0<{\cal T}_1<{\cal T}_2<{\cal T}_3<\cdots$. Moreover, for any $k\geq 0$, ${\cal T}_{k+2}-{\cal T}_k>(\pi/4)(s+c)^{3/2}$. 

{\bf Proof.} For a given elliptic orbit we have $\tau_k+P=\tau_{k+2}$ where $\tau_l$ is the time to go from $\A$ to $\B$ along the arc of type $l$, and $P=2\pi a^{3/2}$ is the Keplerian period, where $a=-(2H)^{-1}$ is the semi-major axis. We write $\inf \tau_k+\inf P< \inf \tau_{k+2}$ where the infimum is on all the elliptic orbits passing through $\A$ and $\B$. The inequality is strict since the variational differential equation (\ref{vde}) excludes $\tau'=H'=0$. This gives the last inequality of the Proposition since according to \citet{gauss}, \S 106, the ellipse of minimal period has  its second focus on the chord $\A \B$, and since ${\cal T}_k=\inf \tau_k$ on the elliptic orbits, in both cases $k\leq 1$ and $k\geq 2$. The inequality ${\cal T}_{2k}<{\cal T}_{2k+1}$ is easily obtained by Lambert's reduction (see 8.1) to the rectilinear case $0<x_\B<x_\A$. Here, starting with the same initial velocity $v_\A$, the arc of type $2k+1$ takes more time than the arc of type $2k$, since when the latter reaches $\B$, the first still have to bounce and to reach $\B$ again. The inequality ${\cal T}_{2k+1}<{\cal T}_{2k+2}$ is obtained by a similar argument. QED

{\bf 2.4. Corollary.} In the planar Kepler problem, for any endpoints $\A$ and $\B$, $\A\neq \B$, for any flight time $\tau>0$, the total number of the Keplerian arcs is finite. It is two when $\tau$ is small enough. It increases with $\tau$ and tends to infinity as $\tau\to +\infty$.

{\bf 2.5. Proposition.} Consider the parabolic arc of type 0 and flight time ${\cal T}_0$, the parabolic arc of type 1 and flight time ${\cal T}_1$ and the unique arc of type $k\geq 2$ and flight time ${\cal T}_k$ introduced in Theorem 2.2. Let ${\cal C}_k$, $k=0,1,2,\dots$ be their respective angular momenta. Then
\begin{equation}\label{T*}
{\cal T}_k=\frac{4(s\Gamma_k^2+2c^2)\Delta}{3(4c^2-\Gamma_k^4){\cal C}_k},\quad\hbox{where}\quad \Gamma_k=\frac{2\Delta}{\sqrt{s^2-c^2}{\cal C}_k}.
\end{equation}

{\bf Proof.} We can check directly this formula in the cases $k=0$ and $k=1$ from (\ref{E*}) and (\ref{C*}). Note that the result of the computation is not so easy to predict. The formula is indeed a surprising consequence of the variational differential equation (\ref{vde}). This formula becomes $3\tau H'=2$ in both cases $k\leq 1$ and $k\geq 2$. It is enough to rationalize this formula by using the angular momentum $C$ as an intermediate variable: $H'=(\d H/\d C)(\d C/\d b)$. QED

{\bf 2.6. Remark.} We will extend Theorem 2.2 to the excluded cases where $\O$, $\A$, $\B$ are collinear. In a first case, which we call the {\it opposition case}, $\O$ is strictly between $\A$ and $\B$. In a second case, which we call the {\it rectilinear case}, $\B$ is strictly between $\O$ and $\A$. The statement may be adapted in order to include each of these cases. However, the adaptation is different in each case. In the opposition case, the arcs of types $2k$ and $2k+1$ are isometric. Indeed, any Keplerian arc may be rotated in space around the axis $\A\O\B$, giving infinitely many arcs. We may still have a statement where even and odd types are replaced by clockwise and anticlockwise types, and where the motions are all restricted to a given plane.
In the rectilinear case, the orbits are rectilinear, the clockwise and anticlockwise planar orbits merge together. The motion may experience collisions, after which it is classically extended by a bouncing which preserves the energy. The number $k$ of half-turns should be adapted and defined as the number of completed periods plus the number of collisions with $\O$. In the more particular cases  $\O=\B$ or $\A=\B$, the questions are simpler, but the statements are different. We restrict the statement of the Theorem to the main case in order to avoid a too long statement. However, the main novelty of this work is the introduction of a variable $b$ which parametrizes the Keplerian arcs in all the cases with $\A\neq \B$. Our method of proof is carefully selected in order to apply to all these cases. This may be useful, since the collinearity of $\O$, $\A$ and $\B$ creates technical problems when the question is to follow a Keplerian arc changing while $\A$ or $\B$ are continuously changing (see \citealt{russell}).

\bigskip

\centerline{\bf 3. Brief description of the arcs from $\A$ to $\B$}

\bigskip

 We always consider that after a collision with $\O$ the Keplerian arcs are continued in the classical way. The velocity tends to infinity when going to collision. The body ``bounces'' and goes back on the same rectilinear path with opposite velocity. The energy is unchanged at the collision. See e.g.\ \citet{alb}.

{\bf 3.1. Definition.} Given two points $\A$ and $\B$, $\A\neq \O$, $\B\neq \O$, we say that the pair of vectors $(p_\A,p_\B)$ is a {\it pair of compatible terminal velocities} or in brief a {\it terminal pair} if the Keplerian orbit around $\O$ which passes at $\A$ with velocity $p_\A$ passes at $\B$ with velocity $p_\B$.

{\bf 3.2. Remark.} If there is an arc starting from $\A$ with velocity $p_\A$ and {\it later} arriving at $\B$ with velocity $p_\B$, then $(p_\A,p_\B)$ is a terminal pair.  As well, if there is an arc starting from $(q_\B,p_\B)$ and later arriving at $(q_\A,p_\A)$, then $(p_\A,p_\B)$ is a terminal pair. Both sorts of arc coexist in the case of an elliptic orbit. Terminal pairs may be continuously changed from a situation where there is a unique hyperbolic arc which is from $\A$ to $\B$ to a situation where there is a unique hyperbolic arc which is from $\B$ to $\A$. These arcs are the two hyperbolic continuations of two complementary elliptic arcs.

The classical first integrals of the planar Kepler problem, where $q=(x,y)\in\R^2$, $p=\dot q=(v,w)\in\R^2$, are the angular momentum $C$, the eccentricity vector $(\alpha,\beta)$ and the energy $H$:
$$C=x w-yv, \quad \alpha= \frac{x}{r}-Cw,\quad \beta=\frac{y}{r}+Cv,\quad H=\frac{v^2+w^2}{2}-\frac{1}{r}.$$
By combining the first three equalities we find the equation of the orbit $r=\alpha x+\beta y+C^2$. Note that if the orbit is hyperbolic only the relevant branch of the hyperbola is represented by this equation.

{\bf 3.3. Proposition.} The two vectors $p_\A$ and $p_\B$ form a terminal pair if and only if the angular momentum, the eccentricity vector and the energy have the same value at $(q_\A,p_\A)$ as at $(q_\B,p_\B)$.

{\bf Proof.} In the nonrectilinear case we assume a common value at $\A$ and $\B$ of $(C,\alpha,\beta)$. It determines the equation of a Keplerian orbit on which at least an arc may be found. So the pair of velocities is a terminal pair. In the rectilinear case, the energy is the only relevant first integral. Starting on the $\O x$ axis at $(x_\A,v_\A)$, the body passes at $x_\B$ with the velocity $v_\B$ or $-v_\B$ due to the energy conservation. If it is $-v_\B$ at a passage, then it is $v_\B$ at another passage. Meanwhile occurs a bouncing at $\O$ or a culmination, which changes the sign of the velocity. QED

{\bf 3.4. Proposition.} We assume that the triangle $\O \A \B$ is not flat.
The map which associates to a terminal pair its angular momentum is one to one. Consequently for any angular momentum $C\in\;]-\infty,0[\;\cup \;]0,+\infty[$, there is a unique terminal pair $(p_\A,p_\B)$ with angular momentum $C$. This pair is a linear combination of $C$ and $C^{-1}$ with coefficients depending on $\O\A\B$.

{\bf Proof.} When fixing a nonzero $C$, Proposition 3.3 gives a linear system of 4 equations in the 4 variables $(v_\A,w_\A,v_\B,w_\B)$:
\begin{equation}\label{syst}
\EQM{x_\A w_\A-y_\A v_\A=C,\quad -w_\A+w_\B&=-C^{-1}r_\A^{-1}x_\A+
C^{-1}r_\B^{-1}x_\B
,\cr
x_\B w_\B-y_\B v_\B=C,\qquad v_\A-v_\B&=-C^{-1}r_\A^{-1}y_\A+
C^{-1}r_\B^{-1}y_\B.}
\end{equation}
The energy integral is omitted, since the energies are equal if these 4 equations are satisfied. The determinant $x_\A y_\B-y_\A x_\B=\Delta$  is nonzero due to the nonflatness hypothesis. QED

{\bf Formula for the terminal velocities.} Let us solve the linear system (\ref{syst}). The right-hand side is sum of a term in $C$ and a term in $C^{-1}$. Thus the solution is sum of a term in $C$ and a term in $C^{-1}$. The term in $C$ satisfies $v_\A=v_\B$ and $w_\A=w_\B$. Then $v_\A=(x_\B-x_\A)C/\Delta$, $w_\A=(y_\B-y_\A)C/\Delta$.
The term in $C^{-1}$ satisfies $(v_\A,w_\A)=\lambda_\A(x_\A,y_\A)$, $(v_\B,w_\B)=\lambda_\B(x_\B,y_\B)$ for some $(\lambda_\A,\lambda_\B)\in \R^2$. Then
$$C\lambda_\A \Delta =r_\B-\frac{x_\A x_\B +y_\A y_\B}{r_\A}=\frac{2r_\A r_\B+c^2 -r_\A^2-r_\B^2}{2r_\A},$$
$$C \lambda_\B \Delta =\frac{x_\A x_\B +y_\A y_\B}{r_\B}-r_\A=\frac{-c^2+r_\A^2+r_\B^2-2r_\A r_\B}{2r_\B}.$$
Heron's formula is
$4\Delta^2=(s^2-c^2)(c^2-(r_\A-r_\B)^2)$ so
$$\frac{C}{\Delta} \lambda_\A=\frac{2}{r_\A(s^2-c^2)},\qquad \frac{C}{\Delta} \lambda_\B=-\frac{2}{r_\B(s^2-c^2)}.$$
The solution is
\begin{equation}\label{p*}\pmatrix{v_\A\cr w_\A\cr v_\B\cr w_\B}=\frac{C}{\Delta}\pmatrix{x_\B-x_\A\cr y_\B-y_\A\cr x_\B-x_\A\cr y_\B-y_\A}+\frac{2\Delta}{(s^2-c^2)C}\pmatrix{r_\A^{-1} x_\A\cr r_\A^{-1} y_\A\cr -r_\B^{-1} x_\B\cr -r_\B^{-1} y_\B}.
\end{equation}
Similar formulas with other expressions of the two coefficients were given by \citet{jacobi}, \citet{jacobif}, p.\ 92, p.\ 96, \citet{godal} or \citet{battin}. Terminal velocities are described in Godal as drawing a hyperbola when $C$ varies. The formula may be interpreted as interpolating the case $C=0^+$, where the arc is in the limit the segment $\B\O$ then the segment $\O\A$,  and the case $C=\infty$, where the arc is the segment $\A\B$.

{\bf 3.5. Corollary.} Assume $\Delta>0$. The real numbers ${\cal C}_0$ and ${\cal C}_1$ given by (\ref{C*}) satisfy $0<-{\cal C}_1<{\cal C}_0$.  The Keplerian arcs of type $k$ have an angular momentum $C\in D_k$ where $D_0=]-{\cal C}_1,+\infty[$, $D_1=]-{\cal C}_0,0[$ and for any $m>1$, $D_{2m}=]-{\cal C}_1,{\cal C}_0[$, $D_{2m+1}=]-{\cal C}_0,{\cal C}_1[$. For any $C\in D_k$ there is  a unique arc of type $k$ and of angular momentum $C$. Its terminal velocities are given by (\ref{p*}).

This Proposition only admits  the arcs with $t_\A<t_\B$. An arc with $t_\A>t_\B$ is obtained from an arc with $t_\A<t_\B$ by changing the signs of the terminal velocities and consequently the sign of $C$. It is natural to consider that for example there are Keplerian arcs with $C\in\;]0,-{\cal C}_1]$. They are of type 1 but have a negative flight time.

Proposition 3.4 defines two branches of terminal pairs, one with $C>0$, the other with $C<0$. When $C>0$ is close to zero, the orbit is hyperbolic, the arc has type 1 and negative flight time. Let us increase $C$.
At $C=|{\cal C}_1|$, the orbit is parabolic. Then, the orbit is elliptic and all the types coexist. At $C={\cal C}_0$, the orbit is parabolic, the arc has type 0 and positive flight time. Finally, for $C>{\cal C}_0$, the orbit is hyperbolic of type 0.

{\bf 3.6. The energy.} Combining $2H=v_\A^2+w_\A^2-2/r_\A$ with (\ref{p*}) gives
$$2H=\frac{C^2c^2}{\Delta^2}+\frac{4(x_\A x_\B+y_\A y_\B-r_\A^2)}{r_\A(s^2-c^2)}-\frac{2}{r_\A}+\frac{4\Delta^2}{(s^2-c^2)^2C^2}$$
or
\begin{equation}\label{H*}2H=\frac{C^2c^2}{\Delta^2}-\frac{4s}{s^2-c^2}+\frac{4\Delta^2}{(s^2-c^2)^2C^2}.
\end{equation}
This fraction factorizes if we use as variables the same square roots $\xi=\sqrt{s+c}$, $\eta=\sqrt{s-c}$ as in Euler's formula (the square root is also related to the $\C \to \C, z\mapsto z^2$ regularizing map, where $\C=\O xy$). We get
\begin{equation}\label{U*}2H\xi^2\eta^2\Gamma^2=(\eta-\xi-\Gamma)(\eta-\xi+\Gamma)(\eta+\xi-\Gamma)(\eta+\xi+\Gamma),\quad\hbox{where }\Gamma=\frac{2\Delta}{\xi\eta C}.\end{equation}
The roots correspond to $C=\pm{\cal C}_0$ and $C=\pm{\cal C}_1$, where  ${\cal C}_0$ and ${\cal C}_1$ are given by formula (\ref{C*}).

The description in Proposition 3.4 has a defect: flat triangles $\O\A\B$ are excluded. Such exclusion should happen due to the following monodromy obstruction. Suppose there is a coordinate determining, as $C$ does for the hyperbolic orbits, a unique Keplerian arc for each triangle, but which would also be defined for a flat triangle. Then, fixing this coordinate and following this arc while $\O$ is turning around $\B$, $\A$ and $\B$ being fixed, the arc would be rectilinear of type 0 at the beginning, but rectilinear of type 1 after one turn. This is a contradiction.

A natural idea while seeing formula (\ref{p*}) is to use $C/\Delta$ as a coordinate instead of $C$. This gives an extension to the rectilinear case, but not to the opposition case, where the denominator $s^2-c^2$ is zero, and where the terminal velocities are not in the line $\B\O\A$. Formula (\ref{U*}) suggests to consider the coordinate $\Gamma$. Here the monodromy obstruction shows that $\sqrt {s-c}$ changes sign when $\O$ crosses the segment $\A\B$.

\bigskip
\centerline{\bf 4. A new coordinate for the families of arcs} 
\bigskip
Consider instead of $C$ or $C/\Delta$ or $\Gamma$ the quantity $b=\langle q_\B,p_\B\rangle-\langle q_\A,p_\A\rangle$, which in the planar case satisfies
\begin{equation}\label{b*}
b=x_\B v_\B+y_\B w_\B-x_\A v_\A -y_\A w_\A=\frac{c^2C}{\Delta}-\frac{2s\Delta}{(s^2-c^2)C}.\end{equation}

The right-hand side, deduced from formula (\ref{p*}), shows that $b$ is an increasing function of $C/\Delta$ in both intervals $]-\infty,0[$ and $]0,+\infty[$. In both intervals $b$ varies from $-\infty$ to $+\infty$. So there are two values of $C$ for each value of $b$ (see Figure 1). Let us see what happens in the cases where the triangle $\O\A\B$ is flat.

{\bf 4.1. Lemma (rectilinear case).} If $\O$, $\B$, $\A$ are on the $\O x$ axis, $0<x_\B<x_\A$, then the motion is rectilinear. The terminal pairs $(v_\A,v_\B)\in\R^2$ are only constrained by the energy equation $v_\A^2-2x_\A^{-1}=v_\B^2-2x_\B^{-1}$. They form the two branches of an equilateral hyperbola, one with $v_\B<0$, the other with $v_\B>0$. Each value of the coordinate $b\in\R$ corresponds to one terminal pair on one branch and one terminal pair on the other branch.

{\bf Proof.} The line $b=x_\B v_\B-x_\A v_\A$ in the plane $(v_\A,v_\B)$ cuts the equilateral hyperbola in two points, one on each branch. QED

{\bf 4.2. Lemma (opposition case).} If $\B$, $\O$, $\A$ are on the $\O x$-axis, $x_\B<0<x_\A$, then all the terminal pairs satisfy $\|q_\A\wedge p_\A\|={\cal C}_0$ where $2{\cal C}_0^{-2}=r_\A^{-1}+r_\B^{-1}$.  The plane of motion is arbitrary. In the plane $\O x y$ where $y_\A=y_\B=0$  we have $v_\A=v_\B$ and this unconstrained real number together with the choice $C=\pm {\cal C}_0$ parametrizes all the terminal pairs. Each value of the coordinate
$b=x_\B v_\B-x_\A v_\A=-(r_\A+r_\B)v_\A$ corresponds to one terminal pair on the branch $C={\cal C}_0$ and one terminal pair on the branch
$C=-{\cal C}_0$.

{\bf Proof.} The third condition (\ref{syst}) is $C^{-1}-w_\A=-C^{-1}-w_\B$ or $C^{-1}-C x_\A^{-1}=-C^{-1}-Cx_\B^{-1}$ or $2C^{-2}=x_\A^{-1}-x_\B^{-1}>0$. The fourth condition (\ref{syst}) is $v_\A=v_\B$. QED

{\bf 4.3. Proposition.} In the plane, if the three points $\O$, $\A$, $\B$ are distinct, then for any $b\in\R$, there are exactly two terminal pairs $(p_\A,p_\B)$ satisfying $\langle q_\B,p_\B\rangle-\langle q_\A,p_\A\rangle=b$.
In the nonrectilinear case, the angular momentum $C$ is positive on one pair, negative on the other pair. In the rectilinear case, the velocity at the inner endpoint is positive on one pair, negative on the other pair. The choice of branch $C/\Delta>0$ in the nonflat case gives the {\it consistent choice} $v_\B<0$ in the rectilinear limit.

{\bf Proof.} In the rectilinear case the coordinate $C/\Delta$ is undetermined and replaced by $(x_\B-x_\A)^{-1}(v_\A+v_\B)/2\in \;]-\infty,0 [\;\cup \;]0,+\infty[$, according to (\ref{p*}). If $C/\Delta>0$, this limit should be positive, and as we know that $|v_\B|>|v_\A|$, $v_\B$ should be negative. QED

 \bigskip
\centerline{\includegraphics[width=50mm]{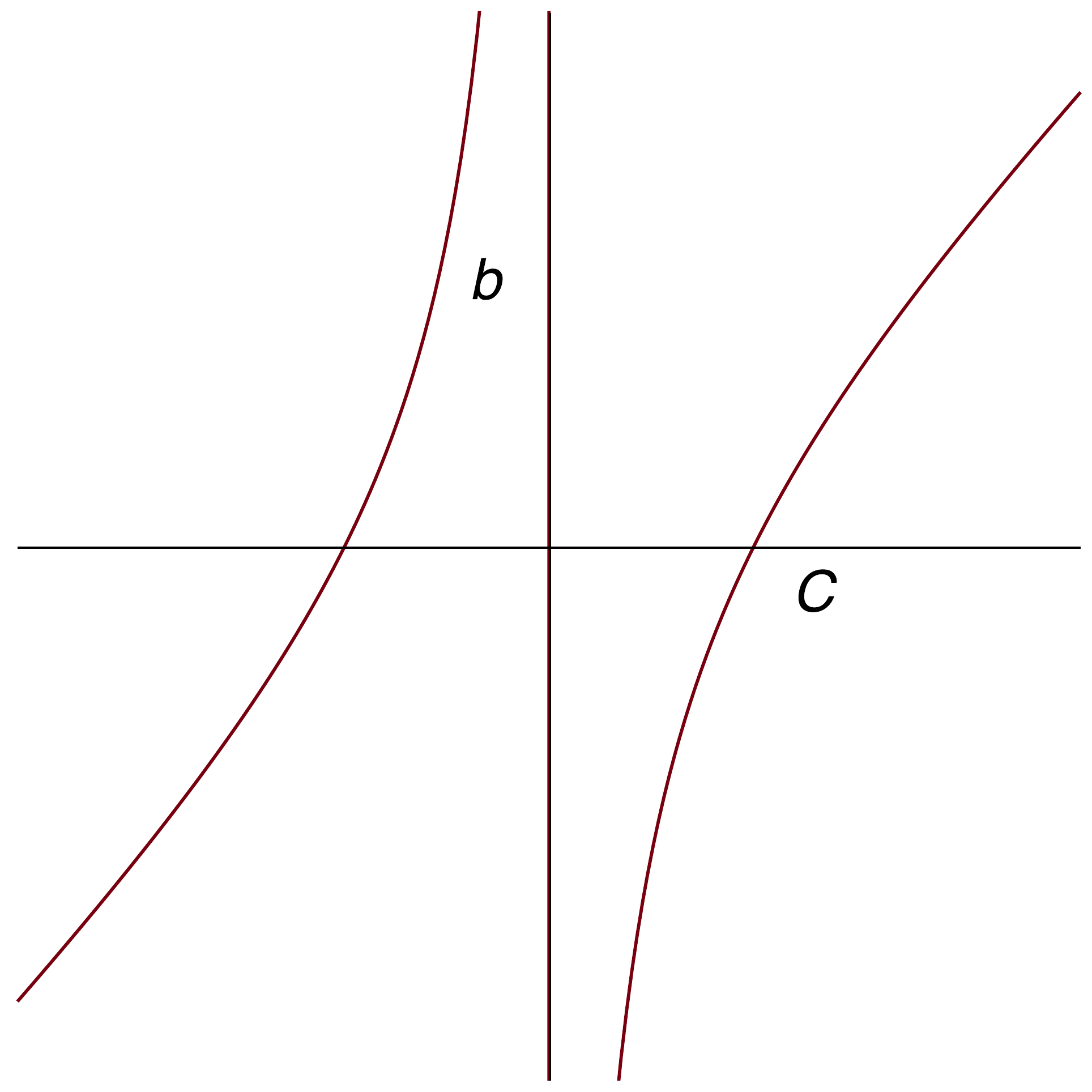}}
\nobreak
\centerline{Figure 1. A graph of $C\mapsto b$ when $\Delta>0$.}
\bigskip

{\bf 4.4. Families of arcs.} The variable $b$ is thus a coordinate for the branches of terminal velocities even in the previously missing cases where $\O$, $\A$, $\B$ are collinear. After restricting it to an interval, $b$ is a coordinate for any family of arcs of Corollary 3.5, but also for the similar families of arcs with collinear $\O$, $\A$ and $\B$. We still need to adapt  to these cases the definition of the type.

{\bf Simplest case.} If $\B=\O$ the motion is rectilinear. Instead of a pair of compatible terminal velocities there is a terminal velocity $v_\A\in\R$. Corresponding to types 0 and 1 there is only one type, with coordinate $v_\A\in\;]-\infty, \sqrt{2/x_\A}[$. For any $l\geq 1$, corresponding to types $2l$ and $2l+1$, there is one type, with coordinate $v_\A\in\;]-\sqrt{2/x_\A}, \sqrt{2/x_\A}[$. The variable $b$ is $-x_\A v_\A$. It is a coordinate of the unique branch.

{\bf Rectilinear case.} The type $k$ of a rectilinear arc starting at $\A$ is the number of times the body bounces off of $\O$, plus the number of times it passes again at $\A$ with the initial velocity $v_\A$. There are arcs of any type $k\in\N$.

{\bf Opposition case.} Here the plane of motion should be fixed. An arc is obtained from another by a reflection of axis $\B\O\A$. There are several ways of defining the type. If we identify an arc with the reflected one, we reduce the number of types as in the simplest case $\B=\O$.

{\bf 4.5. Sign of the energy.} The domain of negative energy in the variable $b$ depends on the choice of branch of terminal pairs. It is $b\in\;]-\xi+\eta,\xi+\eta[$ for a negative $\Gamma\in\;]-\xi-\eta,-\xi+\eta[$, and $b\in\;]-\xi-\eta,\xi-\eta[$ for a positive $\Gamma\in\;]\xi-\eta,\xi+\eta[$. Here again, as in \S 3.6, $\xi=\sqrt{s+c}$ and $\eta=\sqrt{s-c}$. The expressions of the zeros of the energy (the bounds of these domains) in the variable $\Gamma$ and in the variable $b$ are remarkable. Note that the domain in $b$ always contains $b=0$, while the domains in $\Gamma$ never contain $\Gamma=0$. Note that the geometric means of the bounds in $\Gamma$ are respectively $\Gamma=-\sqrt{2c}$ and $\Gamma=\sqrt{2c}$. They both correspond to the minimal energy. Their respective images by (\ref{b*}) are $b=\eta\sqrt{2c}/\xi$ and $b=-\eta\sqrt{2c}/\xi$.

{\bf 4.6. Corollary.} Assume $\A\neq \B$. The quantity $b$ is a coordinate for any of the above described families of arcs of given type. In the multirevolution case, $b\in\;]-\xi+\eta,\xi+\eta[$ or $b\in\;]-\xi-\eta,\xi-\eta[$. For the simple arcs, the interval for $b$ is $]-\infty,\xi+\eta[$ or $]-\xi+\eta,+\infty[$ or $]-\infty,\xi-\eta[$  or $]-\xi-\eta,+\infty[$.

We will now prove that the second derivative with respect to $b$ of the energy $H$ is positive. We should not forget that there are two branches of terminal velocities. To get the value of $H$, we should give $b$ and choose a branch. In the next proposition, the choice should be {\it consistent} in the rectilinear case and in the nonflat case (see Proposition 4.3). If we choose $C/\Delta>0$ in the nonflat case, then  in the rectilinear case $0<x_\B<x_\A$, the choice is $v_\B<0$.

{\bf 4.7. Proposition (algebraic Lambert theorem).} Let $r_\A$, $r_\B$ and $c$ be the three sides of a triangle $\O\A\B$, and $b=\langle q_\B,p_\B\rangle- \langle q_\A,p_\A\rangle$. Let $H(b,r_\A,r_\B,c)$ be the energy of the terminal pair determined by this data and a consistent choice of branch. Then $H(b,r_\A,r_\B,c)=H(b,x_\A,x_\B,x_\A-x_\B)$ where $2x_\A=r_\A+r_\B+c$, $2x_\B=r_\A+r_\B-c$.

{\bf Proof.} The positive value of $C/\Delta$ given by solving (\ref{b*}) only depends on $b$, $s=r_\A+r_\B$ and $c$. 
Now (\ref{H*}) shows that the value of $H$ only depends on $b$, $s$, and $c$.
These three quantities are the same in the left-hand side and in the right-hand side of the proposed equality.  Finally, in the opposition case where $s=c$, the result follows from the formula
$2H={b^2}/{c^2}-{2}/{c}$. QED

{\bf 4.8. Proposition.} Fix a branch of terminal velocities and consider $H: \R\to \R$, $b\mapsto H(b)$. Then $H''(b)>0$. The energy $H$ tends to $+\infty$ when $b\to -\infty$ or $b\to +\infty$. It has a unique minimum, for which $H$ is negative.

{\bf Proof.} According to Proposition 4.7, it is enough to conclude in the rectilinear case, where $0<x_\B<x_\A$, $H=v_\A^2/2+1/x_\A$ and $b=x_\B v_\B-x_\A v_\A$. We have $\d H/\d v_\A=v_\A$ and $\d b/\d v_\A=x_\B v_\A/v_\B-x_\A$ since $v_\A^2/2+1/x_\A=v_\B^2/2+1/x_\B$. Then $$H'=\frac{v_\A v_\B}{x_\B v_\A-x_\A v_\B}=\Bigl(\frac{x_\B}{v_\B}-\frac{x_\A}{v_\A}\Bigr)^{-1},$$
$$H''=\frac{x_\B v_\A^3-x_\A v_\B^3}{(x_\B v_\A-x_\A v_\B)^3}=\Bigl(\frac{x_\B}{v_\B}-\frac{x_\A}{v_\A}\Bigr)^{-3}\Bigl(\frac{x_\B}{v_\B^3}-\frac{x_\A}{v_\A^3}\Bigr)>0.$$
The inequality is deduced from $|v_\B|>|v_\A|$, which also implies that the denominator is nonzero. Now if $b\to\pm\infty$, then $v_\A\to\pm\infty$ or $v_\B\to\pm\infty$, and consequently $H\to +\infty$. Finally, we know that $H$ has two zeros, so the minimum value is negative. QED

\bigskip

\centerline{\bf 5. The variational differential equation}

\bigskip

We chose the coordinate $b$ after the following variational considerations.
Let $q$ be the position and $q\mapsto U$ a force function. In the Kepler problem $U=1/r$. The equation of motion is $\ddot q=\nabla U$. We set $p=\dot q$. The energy $H=\|p\|^2/2-U$ is conserved along the solutions.
We call $(t_\A,q_\A,p_\A)$   the time, the position and velocity vectors at the initial point $\A$, $(t_\B,q_\B,p_\B)$  the same quantities at the final point $\B$. There are several remarkable linear combinations with constant coefficients of the {\it Maupertuis action} and the {\it Levi-Civita time}, i.e.,
$$\int_{t_\A}^{t_\B}\|p\|^2 \d t\quad\hbox{and}\quad \int_{t_\A}^{t_\B}U \d t.$$ In particular, the integral of the Lagrangian function or {\it Hamilton action} from $\A$ to $\B$ is
$$S=\int_{t_\A}^{t_\B}\Bigl(\frac{1}{2}\|p\|^2+U\Bigr) \d t.$$
Hamilton's principle of stationary action gives the laws of dynamics through the variation of $S$ on paths with fixed ends $(t_\A,q_\A)$   and $(t_\B,q_\B)$. A classical formula in \citet{hamilton1}  gives the variation of $S$ during a change of orbit, as a function of the changes of ends $(\delta t_\A,\delta q_\A)$ and $(\delta t_\B,\delta q_\B)$:
$$\delta S=\langle \delta q_\B,\dot q_\B\rangle-\langle \delta q_\A,\dot q_\A\rangle-H(\delta t_\B-\delta t_\A).
$$
A report on the early proofs may be found in \citet{alb}. As well known, Hamilton's ideas were suggested by and extend to other fundamental physical theories. Another combination is
$$\int_{t_\A}^{t_\B} ({1\over 2}\|p\|^2-U) \d t=H(t_\B-t_\A).$$
The function $J=\langle q,\dot q\rangle$ satisfies $\dot J=\|\dot q\|^2+\langle q,\ddot q\rangle=\|p\|^2+\kappa U$ if we assume that $U$ is positively homogeneous of degree $\kappa$, which implies $\langle \nabla U,q\rangle =\kappa U$. This linear combination 
$$\int_{t_\A}^{t_\B} (\|p\|^2+\kappa U) \d t=J_\B-J_\A=b$$
is ``exact'' in the sense that it is a periodic function of $\B$ when $\B$ is moving along a periodic orbit. All these remarkable integrals satisfy remarkable identities. Notably, the Kepler equation
$$(-2H)^{3/2}t=u-e\sin u,$$
appears as such an identity. Here $e$ is the eccentricity, $u$ the  eccentric anomaly. Indeed, differentiating with respect to $t$ gives $(-2H)^{3/2}=(1-e\cos u)\dot u=-2Hr\dot u$. So $(-2H)^{-1/2}u=\int \d t/r$ is the Levi-Civita time. The Kepler equation, when multiplied by $(-2H)^{-1/2}$, is simply
$$-2\int_{t_\A}^{t_\B}({1\over 2}\|p\|^2-U)\d t=\int_{t_\A}^{t_\B}U\d t-\int_{t_\A}^{t_\B}(\|p\|^2-U)\d t,$$
with $t_\A$ fixed at the pericenter.
It is difficult to deduce Theorem 2.2 from the Kepler equation and the usual Keplerian recipe. In contrast, another identity between the above integrals, which includes $S$ instead of the Levi-Civita time, easily gives the proof. We denote by $\tau=t_\B-t_\A$ the signed flight time. Here is the combination.

{\bf 5.1. Lemma.} If the force function $U$ is positively homogeneous of degree $\kappa$, then $(2+\kappa)S=2b+(-2+\kappa)H\tau$. If $\kappa=-1$ as in the case of the Newtonian force function then $S=2b-3H\tau$.

{\bf 5.2. Lemma.} For any natural integer $k$, consider $b$ as parametrizing the arcs of type $k$ from $\A$ to $\B$. The flight time $\tau$ satisfies the ``variational'' differential equation
\begin{equation}\label{vde}
2H\tau'+3\tau H'=2,
\end{equation}
where $H$ and $\tau$ are expressed as functions of $b$ and where $'$ marks the derivative with respect to $b$.

{\bf Proof.} The variation of $S$ with fixed ends is $\delta S=-H\delta\tau$. The variation of the previous identity is $2H\delta\tau+3\tau \delta H=2\delta b$. It is enough to take $b$ as the parameter for the variation. QED

{\bf 5.3. Remark.} The period $P$ of the Keplerian orbits passing through $\A$ and $\B$, expressed as a function of $b$, satisfies $2H P'+3P H'=0$. If $\tau$ is the flight time on an arc of type $k$, $\tau+P$ is the flight time on an arc of type $k+2$ with the same terminal pair. Both flight times satisfy the same differential equation.
The differential equation suggests the following renumbering of the types of arcs. An arc of type $2l$ should have index $l$. An arc of type $2l-1$ should have index $-l$. The flight time $\tau_0$ on an arc of type 0, or index 0, is a solution of the differential equation. Then the {\it signed} flight time on an arc of index $l$ with the same terminal pair is $\tau_0+lP$ and it satisfies the same differential equation.

\bigskip

\centerline{\bf 6. Proof of Theorem 2.2 with the variational differential equation.}

\bigskip

Theorem 2.2 is an easy corollary of this Proposition.

{\bf 6.1. Proposition.} Consider the Keplerian arcs from $\A$ to $\B$, $\A\neq \B$, of a given type as a one parameter family, the parameter being the quantity $b=\langle q_\B,p_\B\rangle-\langle q_\A,p_\A\rangle$. The positive flight time $\tau$ on this arc, considered as a function of $b$, does not have a local maximum. If the type corresponds to simple arcs, $\tau(b)$ does not have a critical point.

{\bf Proof.} The derivative  with respect to $b$ of the variational differential equation $2H\tau' +3\tau H'=2$ is
$$ 2H\tau''+5H'\tau'+3H''\tau=0.$$
Proposition 4.8 concludes that $H''(b)>0$ everywhere. At a critical point $b_c$ we have $\tau'(b_c)=0$ and consequently $H(b_c)\tau''(b_c)<0$. Every critical point in the interval where $H<0$ is a local minimum. In the multirevolution case, we have $H<0$ and the proof is finished.
In the case of a family of simple arcs, any critical point $b_c$ in the interval with $H>0$ should be a local maximum. But at a bound $b_p$ of this interval $H(b_p)=0$ and consequently $H'(b_p)\tau'(b_p)<0$. According to Corollary 4.6 this interval is $]-\infty,b_p[$ or $]b_p,+\infty[$. In the first case, $H'(b_p)<0$ and $\tau'(b_p)>0$. There should be a local minimum in the interval $]b_c,b_p[$. This is a contradiction. In the second case, $H'(b_p)>0$
and there is the same contradiction. To further conclude that there are no critical points in the case of simple arcs, we apply the same argument to the interval with $H<0$. Between a bound $b_p$ of this interval and a local minimum $b_c$, there would be a local maximum, which is impossible. QED

{\bf 6.2. Remark.} It may be considered as inelegant to specify $b$ as the parameter and then state a conclusion which essentially does not depend on the parameter. However, we should consider that parameters are often proposed which are not monotone functions of $b$, such as the energy or the eccentricity.
Such a parameter could in principle produce, at an arc where the monotonicity is lost, a local maximum. About the angular momentum, it is monotone but constant if $\O$, $\A$ and $\B$ are on a line. In brief, the specification of $b$ makes the statement shorter.

\bigskip

\centerline{\bf 7. Convexity of the flight time. Particular results.}

\bigskip

{\bf 7.1. Proposition (arcs finishing at collision).} Let $\eta(x_I,v_I)\in\R^+$ be the flight time  up to the Keplerian collision of a body dropped without angular momentum from $x_I\in\R^+$ with a velocity $v_I\in\R$. The successive derivatives of $\eta$ with respect to $v_I$ are all positive.

{\bf Proof.} We parametrize the free fall by the velocity $v$, which satisfies Newton's law $\dot v=-x^{-2}$. We have
\begin{equation}\label{i*}\eta(x_I,v_I)=\int_{-\infty}^{v_I}x^2\d v,\quad\hbox{with}\quad  x=\frac{2}{v^2-2H}, \quad H=\frac{v_I^2}{2}-\frac{1}{x_I}.\end{equation}
We change the variable $v$ to  $u=v_I-v$. Then
$$\eta(x_I,v_I)=\int_{0}^{+\infty}\hat x^2\d u,\quad\hbox{with}\quad  \hat x=\frac{2}{(v_I-u)^2-2H}.$$
This relation gives $1/\hat x=-uv_I+u^2/2+1/x_I$, and if $u$ is fixed, ${\d \hat x}/{\d v_I}=\hat x^2u$. Denoting by $\eta'$ the partial derivative with respect to $v_I$, we get
$$\eta'=\int_{0}^{+\infty}2x^3 u \d u>0,\qquad \eta''=\int_{0}^{+\infty}6x^4 u^2 \d u>0, \qquad \hbox{etc. QED}$$

{\bf 7.2. Differential equation.} As a limit case of the variational differential equation (\ref{vde}), we have $2H \eta'+3v_I\eta=-2x_I$. We can also obtain this equation by differentiation and integration by parts from (\ref{i*}). In the case $H=0$, the differential equation becomes the identity $3v_I\eta=-2x_I$.

{\bf Remark.} If the motion continues by bouncing a given number of times, still finishing at collision, the flight time is still a convex function of $v_I$, but is no longer increasing. It decreases from infinity, reaches a minimum and increases again to infinity, while $v_I$ varies between the escape velocity and its opposite. At the minimum the differential equation gives again the relation $3v_I\eta=-2x_I$.

Proposition 7.1 will now be extended to arcs from $\A$ to $\B$ where $\B$ is not assumed to coincide with $\O$. The result depends on which coordinate for the arcs replaces $v_\A$.

{\bf 7.3. Proposition (arcs of type 0).} In the plane $\O xy$, the Keplerian arcs of type 0 around $\O$ whose ends $\A$ and $\B$ are distinct, symmetric with respect to the vertical axis $\O y$ and placed at an ordinate $y_\A>0$, are parametrized by the inverse of the angular momentum $\upsilon=1/C$ varying in the interval $]0,(r_\A-y_\A)^{-1/2}[$. The flight time $\tau_0=t_\B-t_\A$ is a function of $\upsilon$ whose successive derivatives are all positive. We have: $\tau_0\to 0$ when $\upsilon\to 0$ and $\tau_0\to +\infty$ when $\upsilon\to (r_\A-y_\A)^{-1/2}$.

{\bf 7.4. Corollary.} In the Euclidean plane or space, consider three points $\O$, $\A$, $\B$ forming a nonflat triangle with $r_\A=r_\B$.
There is a unique Keplerian arc of type 0 around $\O$ going from $\A$ to $\B$ in a given positive flight time. This arc is in the plane $\O\A\B$ and is symmetric with respect to the perpendicular bisector of $\A \B$.

 \bigskip
\centerline{\includegraphics[width=50mm]{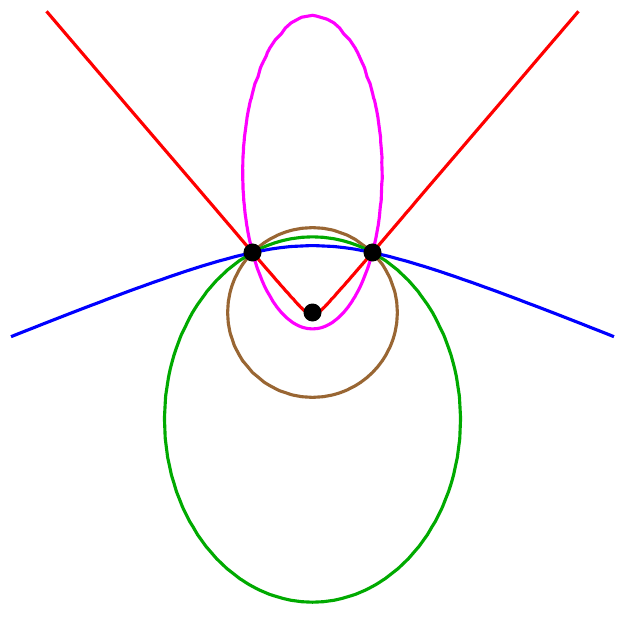}}

\nobreak
\centerline{Figure 2. The family of orbits in the case $r_\A=r_\B$.}
\bigskip

{\bf Proof.} It is easily proved that the arc is symmetric with respect to the $\O y$ axis (see \citealt{AlU}). It thus belongs to a conic section with polar equation
$$
r=\frac{C^2}{1-\beta \sin \theta},
$$
with $C>0$ and $\beta\in \R$. The absolute value $|\beta |$ is the eccentricity, $C$ is the angular momentum, $C^2$ is the semi-parameter. The conic section passes through $\A$, of polar coordinates $(r_\A,\theta_\A)$, with $\theta_\A\in\;]0,\pi/2[$, and through $\B$, of polar coordinates $(r_\B,\theta_\B)=(r_\A,\pi -\theta_\A)$. This is expressed by the single condition
$$
r_\A=\frac{C^2}{1-\beta \sin \theta_\A},\quad\hbox{which is}\quad r_\A-C^2=\beta y_\A
$$
with $\beta\in\;]-\infty, r_\A/y_\A[$ or equivalently with $C^2\in\;]0,+\infty[$. The new form of the polar equation is
$$
r=\frac{C^2}{1-y_\A^{-1}(r_\A-C^2) \sin \theta}=\frac{y_\A}{\sin \theta-(r_\A\sin \theta-y_\A)C^{-2}}.
$$
At the ends of the interval, the arc is a limit of hyperbolas. When $\beta\to -\infty$, it is a segment going from $\A$ to $\B$. When $\beta\to \beta_0$, it is a pair of segments, from $\A$ to $\O$ and then from $\O$ to $\B$ (see Figure 2). The arc of type 0 is the upper arc, which exists if and only if $\beta\in\;]-\infty,1[$. This is $C^2\in\;]r_\A-y_\A,+\infty[$. We express the flight time along the symmetric arc of type 0 by using the expression $C=r^2\dot\theta$ of the angular momentum:
$$
\tau_0=\int_{t_\A}^{t_\B} \d t=\int_{\theta_\A}^{\theta_\B}\frac{r^2}{C}\d\theta=y_\A^2C^{-1}\int_{\theta_\A}^{\theta_\B}
\bigl( \sin \theta-(r_\A\sin \theta-y_\A)C^{-2}\bigr)^{-2}\d\theta.$$
It is natural to use the variable $\upsilon=C^{-1}\in\;]0,(r_\A-y_\A)^{-1/2}[$.
We differentiate under the integration symbol. Note that the function
$$
{r^2\over y_\A^2C}=\frac{\upsilon}{\bigl(\sin \theta-(r_\A\sin \theta-y_\A)\upsilon^2\bigr)^2}=\frac{\upsilon}{(h-g \upsilon^2)^2}
$$
with $h=\sin\theta>0$ and $g=r_\A\sin \theta-y_\A>0$ is already the derivative in $\upsilon$ of
$$
{r\over 2y_\A g}=\frac{1}{2g(h-g \upsilon^2)}.
$$
The function $(h-g\upsilon^2)^{-1}$ with $h>0$ and $g>0$ has all its successive derivatives in $\upsilon$ positive in the domain $\upsilon\in\;]0,\sqrt{h/g}[$. Consequently $\d^n(\tau_0(\upsilon))/\d \upsilon^n>0$ for any $n\in\N$. QED

\bigskip

{\bf The arcs of type 1 in the rectilinear case.} We use the notation $\eta$ for the flight time to collision as in Proposition 7.1 and furthermore set $\eta_\A(v)=\eta(x_\A,v)$, $\eta_\B(v)=\eta(x_\B,v)$. The type 1 flight time is
$\tau_1=\eta_\A(v_\A)+\eta_\B(-v_\B)$ since after bouncing we arrive at $\B$ with a $v_\B>0$, while $\eta$ was counted in Proposition 7.1 from $\B$ to $\O$, starting with the opposite velocity.
We compute the first and second derivatives of $\tau_1$
with respect to $b=x_\B v_\B-x_\A v_\A$ considered as the coordinate on the branch $v_\B>0$ of terminal pairs $(v_\A,v_\B)\in \R^2$ satisfying $v_\A^2-2x_\A^{-1}=v_\B^2-2x_\B^{-1}$. The positions $x_\A$ and $x_\B$ are fixed and satisfy $0<x_\B<x_\A$. We have
$$\tau_1'=\eta_\A'(v_\A)v_\A'-\eta_\B'(-v_\B)v_\B',\qquad \tau_1''=\eta_\A''(v_\A)v_\A'^2+\eta_\B''(-v_\B)v_\B'^2+R,$$
$$R=\eta_\A'(v_\A)v_\A''-\eta_\B'(-v_\B)v_\B''.$$
According to Proposition 7.1, the first two terms of $\tau_1''$ are always positive. We will prove that $R>0$. As $0=b''=x_\B v_\B''-x_\A v_\A''$, there is a function $\lambda$ such that $v_\A''=\lambda x_\A^{-1}$, $v_\B''=\lambda x_\B^{-1}$. Let us compute $\lambda$.
The energy relation is  $v_\A v_\A'=v_\B v_\B'$.
 We have $\d (x_\B v_\B-x_\A v_\A)/\d v_\A=(x_\B v_\A-x_\A v_\B)/v_\B$ and
$$v_\A'=\frac{v_\B}{x_\B v_\A-x_\A v_\B},\qquad v_\A''=v_\A'\frac{x_\B(v_\A^2-v_\B^2)}{v_\B(x_\B v_\A-x_\A v_\B)^2}=\frac{x_\B(v_\A^2-v_\B^2)}{(x_\B v_\A-x_\A v_\B)^3}.$$
$$\lambda=\frac{x_\A x_\B(v_\A^2-v_\B^2)}{(x_\B v_\A-x_\A v_\B)^3}=\frac{x_\B-x_\A}{(x_\B v_\A-x_\A v_\B)^3},\qquad R=\lambda\Bigl(\frac{\eta_\A'(v_\A)}{x_\A}-\frac{\eta_\B'(-v_\B)}{x_\B}\Bigr).$$
We see that $\lambda$ has the sign of $v_\B$, since $-x_\A v_\B$ dominates in the denominator and the numerator is negative. So $\lambda$ is positive and we are left to prove that the parenthesis is positive.

{\bf 7.5. Lemma.} Let a particle be ejected from the Keplerian collision with a center $\O$, at time $t=0$. Call $v\in\R$ its velocity, $x>0$ its position, $H=v^2/2-1/x$ its energy. Then $tv/x\to 2/3$ as $t\to 0$. On all the motion in positive time, $tv/x$ is decreasing up to $-\infty$ if $H<0$, constant if $H=0$, increasing up to $+\infty$ if $H>0$.

{\bf Proof.} In the case $H=0$, $x=gt^{2/3}$, $v=2gt^{-1/3}/3$ with $g=\root 3 \of {9/2}$. This gives $tv/x=2/3$, and also gives the limiting value for $H\neq 0$. If $H>0$, the Keplerian recipe is $a=-(2H)^{-1}$, $x=a(1-\cosh u)$, $-|a|^{-3/2}t=u-\sinh u$, $\d u/\d t=(\cosh u-1)^{-1}|a|^{-3/2}$, $v=|a|^{-1/2}\sinh u(\cosh u-1)^{-1}$. Thus
$$\frac{tv}{x}=\frac{\sinh^2 u}{(\cosh u-1)^2}-u\frac{\sinh u}{(\cosh u-1)^2},$$
$$\frac{\d}{\d u}\frac{tv}{x}=-\frac{3\sinh u}{(\cosh u-1)^2}+u\frac{2+\cosh u}{(\cosh u-1)^2}>0\quad\hbox{for all } u>0$$
since the Taylor expansion of the numerator $u^5(1/24-1/40)+u^7(1/720-1/1680)+\cdots$ has no negative terms. If $H<0$, $x=a(1-\cos u)$, $a^{-3/2}t=u-\sin u$, $\d u/\d t=(1-\cos u)^{-1}a^{-3/2}$, $v=a^{-1/2}\sin u(1-\cos u)^{-1}$ thus
$$\frac{tv}{x}=-\frac{\sin^2 u}{(1-\cos u)^2}+u\frac{\sin u}{(1-\cos u)^2},$$
$$\frac{\d}{\d u}\frac{tv}{x}=\frac{3\sin u}{(1-\cos u)^2}-u\frac{2+\cos u}{(1-\cos u)^2}<0 \quad\hbox{for all } u\in\;]0,2\pi[$$
since $3\sin u/(2+\cos u)-u$ is zero
at $u=0$ and its derivative $-(1-\cos u)^2/(2+\cos u)^2$ is negative.   QED

{\bf 7.6. Lemma.} Call $\eta'(x,v)$ the partial derivative with respect to $v$, with fixed $x$, of the flight time $\eta(x,v)$ needed to reach the collision by starting from $x$ with velocity $v$. Let us choose an arbitrary energy $H$. Let $x(v)=(v^2/2-H)^{-1}$ be the value of $x$ corresponding to this choice. Then $\eta'(x(v),v)/x(v)$ is increasing with $v$.

{\bf Proof.} According to Proposition 7.1, $\eta'>0$. According to the differential equation $2H\eta'+3v\eta=-2x$,  $\eta'/x=(2H)^{-1}(-2-3v\eta/x)$. If $H\neq 0$, the question is reduced to the monotonicity of the function $v\eta(x(v),v)/x(v)$. But $\eta(x,v)$ is the function $t(x,-v)$ of Lemma 7.5, where the velocity $-v$ is decreasing all along the orbit, and where $H$ is constant all along the orbit. So $v\eta/x$ is increasing with $v$ if $H<0$, decreasing when $v$ is increasing if $H>0$.
Consequently $\eta'/x$ is increasing with $v$ in both cases. If $H=0$, the differential equation gives $v\eta/x=-2/3$, which we already know from Proposition 7.1. Differentiating it, we get $2H\eta''+5v\eta'+3\eta=0$, which gives $5\eta'/x=-3\eta/(xv)=2/v^2$ if $H=0$, which is also increasing with $v$ since $v<0$.
QED

{\bf 7.7. Proposition (arcs of type 1).} In the rectilinear case, $0<x_\B<x_\A$, the second derivative $\tau_1''$ with respect to $b$ of the flight time $\tau_1>0$ is positive. The first derivative $\tau_1'$ is negative.

{\bf Proof.} Since $v_\B>0$ and $v_\B^2-v_\A^2>0$, we have $v_\A>-v_\B$ and according to Lemma 7.6, the above quantity $R$ is positive. This proves $\tau_1''>0$. Now as $v_\A\to -\infty$, $\tau_1\to 0$, $v_\B\to +\infty$ and $b\to +\infty$. As we proved that $\tau_1'$ decreases, $\tau_1'<0$. QED 

\bigskip

\centerline{\bf 8. Convexity of $\tau$. General results and second proof of Theorem 2.2.}

\bigskip

{\bf 8.1. Lambert's theorem and L-congruence.} According to Lambert's theorem (see \citealt{lambert,gooding,alb}), by moving $\O$ on an ellipse with foci $\A$ and $\B$, we change continuously all the Keplerian arcs of type $k$ going from $\A$ to $\B$ in a given flight time $\tau_k$, without changing either their number or the integer $k$ or the energy of each arc. \citet{gooding} calls L-congruents the arcs obtained from each other by following such a continuous change (see Figure 3).

\vspace{0.5cm}
\centerline{\includegraphics[width=65mm]{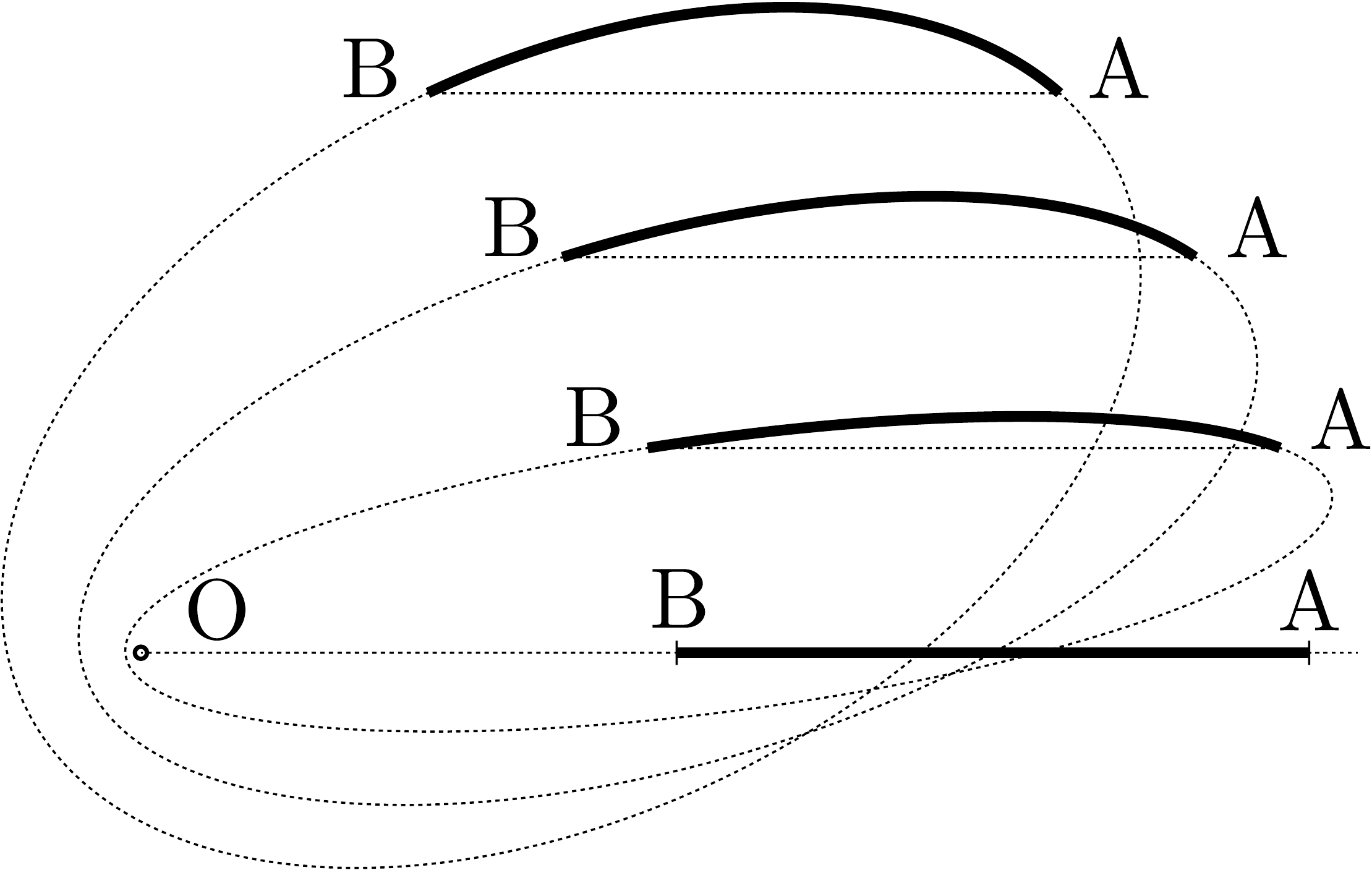}}
\centerline{Figure 3. Four Keplerian arcs of type 0 with same $c$, $r_\A+r_\B$ and $H$.}
\vspace{0.5cm}

The L-congruence furthermore leaves invariant the quotient $C/\Delta$, as shown by formula (\ref{H*}). Formula (\ref{b*}) shows that $b=\langle q_\B,p_\B\rangle-\langle q_\A,p_\A\rangle$ is also invariant. We gave in 4.7 an ``algebraic Lambert theorem'' which indeed states the L-invariance of $b$. The variational differential equation is L-invariant, which indicates an alternative proof of the L-invariance of $\tau$, which is Lambert's theorem.

The L-congruence reduces Lambert's problem to the case with $r_\A=r_\B$. As well, it reduces it to the case where $\O$, $\B$, $\A$ are on a line in this order. So, Proposition 7.3 extends to the nonsymmetric cases. The opposition case should nevertheless be excluded. In the same way, Proposition 7.7 extends to the nonrectilinear cases, including the opposition cases which are L-congruent to the case of Proposition 7.1.

{\bf 8.2. Convex composition of functions.} If the (smooth) real function $\varphi$ satisfies $\varphi'>0$ and $\varphi''>0$, if the real function $f$  satisfies $f'>0$ and $f''\geq 0$, then the composed function $g: y\mapsto f(\varphi(y))$  satisfies $g'>0$ and $g''>0$. This is proved by the formulas $g'=\varphi' f'$ and $g''=\varphi'' f'+(\varphi')^2f''$. The same formulas prove the convexity of $g$ in other cases.

Here, we restrict the discussion to the coordinates $\Delta/C$ and $b$, which are related by (\ref{b*}). For other coordinates and references see \citet{AlU}.

Proposition 7.3 proves that the flight time $\tau_0>0$ satisfies $\tau_0'>0$ and $\tau_0''>0$. Here $'$ is the derivative with respect to the coordinate $\Delta/C$. Here $C>0$ and $\Delta>0$ since $x_\A=-x_\B>0$ and $y_\A=y_\B>0$. Let us pass to the coordinate $b$.

The change of variable $\Delta/C\mapsto b$ given by (\ref{b*}) satisfies $b'<0$. If $C/\Delta <0$, $b''<0$. If $C/\Delta>0$, $b''>0$. We are in this last case. We need the reciprocal change: $\varphi: b\mapsto \Delta/C$. We have $\varphi'<0$, $\varphi''>0$.
Consequently the function $b\mapsto \tau_0$ has a negative first derivative and a positive second derivative.

Proposition 7.7 proves that the flight time $\tau_1>0$ satisfies $\tau_1'<0$ and $\tau_1''>0$. Here~$'$~is the derivative with respect to the coordinate $b$. Here $v_\B>0$ which is consistent with $\Delta/C<0$. Let us pass to the coordinate $\Delta/C$.
Here $\varphi(\Delta/C)=b$, $\varphi'<0$, $\varphi''<0$. Then $C/\Delta\mapsto \tau_1$ has a positive first derivative and a positive second derivative.

{\bf 8.3. Theorem.} Consider a nonflat triangle $\O\A\B$. Consider the Keplerian arcs from $\A$ to $\B$ of any given type as a one parameter family, the parameter being $1/C$, inverse of the angular momentum. The positive flight time $\tau$ on this arc, considered as a function of $1/C$, has a positive second derivative.

{\bf Proof.} About $\tau_0$, we extend Proposition 7.3 to general arcs by L-congruence. About $\tau_1$, we extend Proposition 7.7 by L-congruence and pass to the variable $1/C$ by convex composition. So both $\tau_0$ and $\tau_1$ have a positive second derivative in $1/C$. We obtain a positive second derivative of the Keplerian period by adding $\tau_0(-1/C)$ and $\tau_1(1/C)$. By similar additions, we include the multirevolution flight times. QED

This Theorem extends to the rectilinear case by changing the coordinate $1/C$ into $v_\A-v_\B$, as shown by computing $v_\A-v_\B$ with formula (\ref{p*}). Note that $(v_\A+v_\B)(v_\A-v_\B)=2/x_\A-2/x_\B$ since the energy in $\A$ is the energy in $\B$. The quantities $v_\A$, $H$, $b$ are rational functions of $v_\A-v_\B$. But Theorem 8.3 cannot be extended to the opposition case. This is proved by the monodromy obstruction explained at the end of \S 3. In the next Theorem, all the cases are included.

{\bf 8.4. Theorem.} Consider the Keplerian arcs from $\A$ to $\B$, $\A\neq \B$, of any given type as a one parameter family, the parameter being the quantity $b=\langle q_\B,p_\B\rangle-\langle q_\A,p_\A\rangle$. The positive flight time $\tau$ on this arc, considered as a function of $b$, has a positive second derivative.

{\bf Proof.} About $\tau_0$, we extend Proposition 7.3 to general arcs by L-congruence. The rectilinear case is included by changing $\Delta/C$ into $v_\A-v_\B$. We then pass to the variable $b$ by convex composition. The opposition case is included by extending Proposition 7.1 by L-congruence. About $\tau_1$, we extend Proposition 7.7 by L-congruence. We conclude as in the previous Theorem.
 QED

Theorem 8.4 gives the convexity of $\tau$, while Proposition 6.1 only proves that $\tau$ does not have a local maximum. Our main Theorem 2.2 was deduced from Proposition 6.1. It is {\it a fortiori} deduced from Theorem 8.4.

\bigskip

{\bf Acknowledgements.} We wish to thank Prof.\ Giovanni Valsecchi for the reference
\citet{henon}. We thanks both reviewers for helping us to improve our work.



\end{document}